\documentclass[aps,pra,twocolumn,showpacs]{revtex4}
\usepackage{graphicx,amsmath}

\newcommand{\er}{\mathbf{r}}
\newcommand{\de}{\mathrm{d}}
\newcommand{\ee}{\mathrm{e}}

\begin{document}

\title{Engineering optical soliton bistability in colloidal media}

\author{Micha\l{} Matuszewski}

\affiliation{Nonlinear Physics Center, Research School of Physics and Engineering, Australian National
University, Canberra ACT 0200, Australia}

\begin{abstract}
We consider a mixture consisting of two species of spherical nanoparticles dispersed in a liquid medium.
We show that with an appropriate choice of refractive indices and particle diameters, 
it is possible to observe the phenomenon of optical soliton bistability 
in two spatial dimensions in a broad beam power range. Previously, this possibility was ruled out in the case of a single-species colloid. 
As a particular example, we consider the system of hydrophilic silica particles and gas bubbles generated in the process
of electrolysis in water.
The interaction of two soliton beams can lead to switching of the lower branch solitons to the upper branch,
and the interaction of solitons from different branches is phase-independent and always repulsive.

\end{abstract}
\maketitle

\section{Introduction}

Spatial optical solitons are formed when the change of nonlinear refractive index induces an effective lensing effect
that balances diffraction of the laser beam~\cite{Book}. When a laser beam passes through a colloidal medium
composed of liquid suspension of dielectric nanoparticles, the optical gradient force acts against particle diffusion,
increasing the refractive index in the regions of higher light intensity. The corresponding local change of the refractive index
is of the self-focusing type, and allows for creation of spatial optical solitons in the form of self-trapped beams, as
was demonstrated in both theoretical and experimental studies~\cite{Ashkin1,Gordon,Segev,Segev_OL,colloidsOE,Dholakia_OE}. 
We note that optical solitons have been also observed in other soft matter systems, including liquid crystals~\cite{Asanto} and polymers~\cite{Pispas}.

Recently, it was shown theoretically~\cite{colloidsOE} that the optical response of a colloidal medium in the hard-sphere approximation
allows for the existence of two different stable soliton solutions for the same beam power, i.e.~the soliton bistability of the first
kind~\cite{Kaplan,secKind}. Moreover, interactions of these bistable solitons have interesting properties, not found in other optical
soliton systems~\cite{colloidsPRA}. Soliton switching through collisions and phase-independent repulsive interactions were demonstrated.
The soliton bistability, however, has been only predicted in the one-dimensional case \cite{colloidsOE}, corresponding to a planar waveguide setup
or surface waves. In contrast, it was shown that in the two-dimensional case the lower soliton branch becomes unstable~\cite{PLP} due to the 
intrinsic instability of multidimensional solitons in Kerr media. Moreover, even in the one dimensional case the power range 
for which bistable solitons exist is limited.

In this paper, we demonstrate that the introduction of a second colloidal species 
allows for the bistability of two-dimensional soliton beams. The idea of ``engineering'' the nonlinear response of a colloidal medium by mixing
several components was first proposed in~\cite{Segev}.
We show that with an appropriate choice of refractive indices and particle dimensions, it is possible to observe the bistability in a broad beam power range. 
As a particular example, we consider the system of hydrophilic silica particles and gas bubbles generated in the process
of electrolysis in water.
In analogy with the one-dimensional case, the interaction of soliton beams can lead to switching to the upper soliton branch,
and the interaction of solitons from different branches is phase-independent and always repulsive.

\section{Model}

We consider a mixture consisting of two species of colloidal particles dispersed in a background liquid medium.
We assume that the first (second) group of particles is characterized by a refractive index higher (lower) than the background medium index.
For the sake of clarity, we will refer to the two groups as to ``particles'' and ``bubbles'', although each of the species can have the form
of solid state, liquid, or gas, as long as the condition imposed on their refractive indices is fulfilled. 
In particular, bubbles can be replaced eg.~by a solution of colloidal microshells \cite{GlassSpheres} to enhance the system stability.
We also assume purely dielectric response
of all the components to the laser light and negligible absorption.

The optical gradient force induced by a laser beam will have the effect of locally increasing the concentration of particles and decreasing
the concentration of bubbles, as was shown previously~\cite{Segev}.
If the background concentration of bubbles (that is, concentration in the absence of laser light)
 is low enough to describe them as a system of noninteracting particles (ideal gas),
the same description will hold in the presence of an optical beam. On the other hand, concentration of particles will increase,
and interactions between them can become important, even if they were negligible at the outset.
We take into account interactions between the particles
in the hard sphere approximation.
We also assume that the (otherwise negligible) interaction between particles and bubbles is repulsive, which ensures stability of the system.

The model of nonlinear laser beam propagation in a colloidal suspension in the hard sphere approximation
was elaborated in \cite{colloidsOE}. In the recent experiment \cite{Dholakia_OE} it was shown that the optical response of colloidal suspensions
of polystyrene beads can be substantially different from that predicted by this approximation. We notice, however, that other types of 
colloidal systems can be reasonably well described by the hard-sphere model \cite{HardSphere}. Moreover, it is plausible to expect that 
the phenomenon of soliton bistability can be observable also in the case of ``soft'' interaction. The necessary requirement here is the saturation of 
colloidal particle concentration at high packing fractions, which must occur due to limited available volume. 
We can therefore treat the hard-sphere approximation
as the first step to describing and understanding the physics of more complicated systems.

In our case, the refractive index of colloidal particles $n_{\rm p}$, bubbles $n_{\rm b}$
and the background index $n_{\rm B}$ fulfill the condition $n_{\rm p}>n_{\rm B}>n_{\rm b}$. 
This assumption is necessary for observation of the phenomena described below.
We also assume that the particle and bubble diameters are much smaller
than the laser wavelength in the background medium, $d_{\rm p}, d_{\rm b} \ll \lambda_0/n_{\rm B}$ (Rayleigh regime).
The osmotic pressure can be calculated from the equation of state \cite{SimpleLiquids}
\begin{equation} \label{state}
\frac{\beta \Pi_\nu}{\rho_\nu} = Z_\nu(\eta_\nu)\,,\qquad \nu = {\rm p}, {\rm b}
\end{equation}
where the index $\nu$ is replaced by ${\rm p}$ for solid particles or ${\rm b}$ for bubbles,
$\beta=1/k_{\rm B} T$, $\Pi_\nu$ is the osmotic pressure, $\rho_\nu$ is the
colloidal particle (or bubble) concentration, $Z_\nu(\eta)$ is the compressibility, and
$\eta_\nu=\rho_\nu V_\nu$ is the packing fraction, where $V_\nu$ denotes the particle volume. 
For bubbles we take the ideal gas compressibility $Z_{\rm b}(\eta_{\rm b})=1$. 
For solid particles interacting through a hard sphere potential, the Carnahan-Starling formula
$Z_{\rm p}(\eta_{\rm p}) \approx (1+\eta_{\rm p}+\eta_{\rm p}^2-\eta_{\rm p}^3)/(1-\eta_{\rm p})^3$ gives a very good
approximation up to the fluid-solid transition at $\eta_{\rm p} \approx 0.5$
\cite{SimpleLiquids}. This phenomenological formula is in agreement
with exact perturbation theory calculations as well as molecular
dynamics simulations.

We assume for the time being that the gradient of the concentration of colloidal particles or bubbles
$\rho(\er)$ is locally parallel to $\hat{x}$, and 
consider a small box of volume $\de V=\de x \de S$, with length $\de
x$ and normal surface $\de S$. The difference in the osmotic pressure exerted on
the right and left surface $\de \Pi$ gives rise to an effective force
acting on the colloidal particles $F_{\rm int}$. It is equal to the
external force that is necessary to sustain the concentration gradient,
and $\de \Pi = -F_{\rm int}/\de S = -f_{\rm int} \rho \de V/\de S =
-f_{\rm int} \rho \de x$, where $f_{\rm int}$ is the average force
acting on a single particle. Using Eq.~(\ref{state}) we get $\de
(\rho Z)/\de x = -f_{\rm int} \rho \beta$. The particle current
density is equal to
\begin{equation}
\overrightarrow{j} = \rho \mu (\overrightarrow{f}_{\rm ex} +
\overrightarrow{f}_{\rm int}) = \rho \mu \overrightarrow{f}_{\rm ex}
- D \nabla (\rho Z)\,,
\end{equation}
where $\mu$ is the particle mobility, and $D=\mu/\beta$ is the
diffusion constant. In the ideal gas limit, this equation becomes
Eq.~(3) of \cite{Segev}. 

Let $m=n_\nu/n_{\rm B}$ be the ratio of
the colloidal particle (or bubble) refractive index to the background refractive
index. Polarizability of a sphere is given by
\begin{equation}
\alpha_\nu= 3 V_\nu \varepsilon_0 n_{\rm B}^2 \delta_\nu,
\end{equation}
where $V_{\nu}=(\pi/6) d_\nu^3$ is the sphere volume and $\delta_\nu=(m_\nu^2 - 1) / (m_\nu^2 + 2)$. If we look for the steady
state ($\overrightarrow{j_\nu}=0$) in the presence of optical field
gradient ($\overrightarrow{f}_{\rm ex}=(\alpha_\nu/4) \nabla |E|^2$) we
obtain
\begin{equation}
\rho_\nu\frac{\alpha_\nu \beta}{4} \frac{\de |E|^2}{\de x} = \frac{\de (\rho_\nu Z_\nu)}{\de x}\,,
\end{equation}
which can be solved analytically to give the dependence between $|E|^2$ and the packing fractions $\eta_\nu$
\begin{equation} \label{Ig}
\frac{\beta}{4} |E|^2= \frac{1}{\alpha_{\rm p}}\left[g(\eta_{\rm p}) - g(\eta_{0p})\right] = \frac{1}{\alpha_{\rm b}}\ln \left(\frac{\eta_{\rm b}}{\eta_{0b}}\right)\,,
\end{equation}
where $g(\eta)=(3-\eta)/(1-\eta)^3+\ln \eta$, and $\eta_{0 \nu}$ is the
background packing fraction of the $\nu$ component.

Assuming relatively low packing fractions, the
corresponding nonlinear refractive index change can be approximately
calculated using the Maxwell--Garnett formula \cite{Garnett}.
For low refractive index contrast ($n_{\nu}/n_{\rm B} \approx 1$) we have
\begin{equation}
n^2_{\rm eff} = \varepsilon_{\rm eff} \approx \varepsilon_{\rm B}\left(1 + 3 \sum_\nu\delta_\nu \eta_\nu\right)  \,.
\end{equation}
Substituting
this formula to the Helmholtz equation $\nabla^2 E + k_0^2 n^2_{\rm
eff} E = 0$, we obtain the propagation equation for the slowly varying
envelope of electric field $u(\tilde{\er})$ defined by
$E(\tilde{\er}) = (2/\sqrt{\beta}) u(\tilde{\er})
\exp\left(i k_0 n_0 \tilde{z}\right)$, where $n_0=n_{\rm B}\left(1 + 3 \sum_\nu \delta_\nu \eta_{0\nu} \right)^{1/2}$
\begin{equation} \label{PE_physical}
i \frac{\partial u}{\partial \tilde{z}} + \frac{1}{2 k_0 n_0}
\left[ \nabla^2_{\tilde{\perp}} u + 3 k^2 \sum_\nu \delta_\nu (\eta_\nu - \eta_{0\nu}) u \right] + \frac{i}{2} \sum_\nu \gamma_\nu u = 0\,,
\end{equation}
where $k=k_0 n_{\rm B} = 2\pi n_{\rm B}/\lambda_0$ and the additional last term on the left hand side accounts for
damping due to Rayleigh scattering from the dielectric spheres.
The damping coefficients are given by $\gamma_\nu= 2 \pi^5 \rho_\nu \delta_\nu^2 d_\nu^6 / (3 \lambda^4)$
\cite{Hulst}, where $\lambda=\lambda_0/n_{\rm B}$.
Additionally, for steady state solutions, relation (\ref{Ig}) gives
\begin{align} \label{ueta}
\alpha_{\rm p} |u|^2&=  g(\eta_{\rm p}) - g(\eta_{\rm 0p})\,, \\
\alpha_{\rm b} |u|^2&=  \ln (\eta_{\rm b} / \eta_{\rm 0b})\, \nonumber
\end{align}
at each point of space. 

We renormalize spatial coordinates according to $\left(\tilde{x},\tilde{y}\right)=  (2/3 k^2)^{1/2} \times (x,y) $ and
$\tilde{z}= \left(2n_0/3 k n_{\rm B}\right)  \times z$, obtaining
\begin{equation} \label{PE}
i \frac{\partial u}{\partial z} + \frac{1}{2} \nabla^2_{\perp} u +  \sum_\nu \delta_\nu (\eta_\nu - \eta_{0\nu}) u + \frac{i}{2} \sum_\nu \Gamma_\nu \eta_\nu u = 0\,,
\end{equation}
where the renormalized damping coefficients are
\begin{equation} \label{damping}
\Gamma_\nu= \frac{4}{3} \pi^3\sqrt{1 + 3 \sum_\nu \delta_\nu \eta_{0\nu}} \left(\frac{d_\nu}{\lambda}\right)^3 \delta_\nu^2\,.
\end{equation}
From Eq.~(\ref{PE}) and formula~(\ref{damping}) we conclude that the effect of scattering losses depends strongly on
the ratio of the particle size to the laser wavelength.

In the following, we consider a particular system of hydrophilic silica particles and hydrogen or oxygen nanobubbles
generated in the process of electrolysis in water \cite{electrolysis}. 
Nanobubbles solutions generated in this way can remain stable for several days.
The interaction between silica particles and bubbles
is repulsive independently of the distance \cite{ParticleBubble}, which ensures the system stability. 
We assume that silica particles have diameter $d_{\rm p}=50$ nm and $n_{\rm p}=1.45$ and bubbles 
have diameter $d_{\rm b}=100$ nm and $n_{\rm b}=1$, while the water refractive index is $n_{\rm B}=1.33$. 
The background concentrations of
particles and bubbles in regions of low light intensity are taken as $\eta_{0 \rm p}=10^{-3}$ and $\eta_{0 \rm b}=10^{-2}$, 
which is consistent with the experiments \cite{electrolysis}.
In real systems, it is not possible to prepare monodisperse solutions. We assume that the dispersion of sizes
is relatively low, which allows us to use the present model with $d_{\rm p}$ and  $d_{\rm b}$ equal to average or effective sizes.
The relatively long laser wavelength $\lambda_0=1064\,$nm allows for lower Rayleigh scattering losses.

\begin{figure}
\begin{center}
  \includegraphics[width=8.5cm]{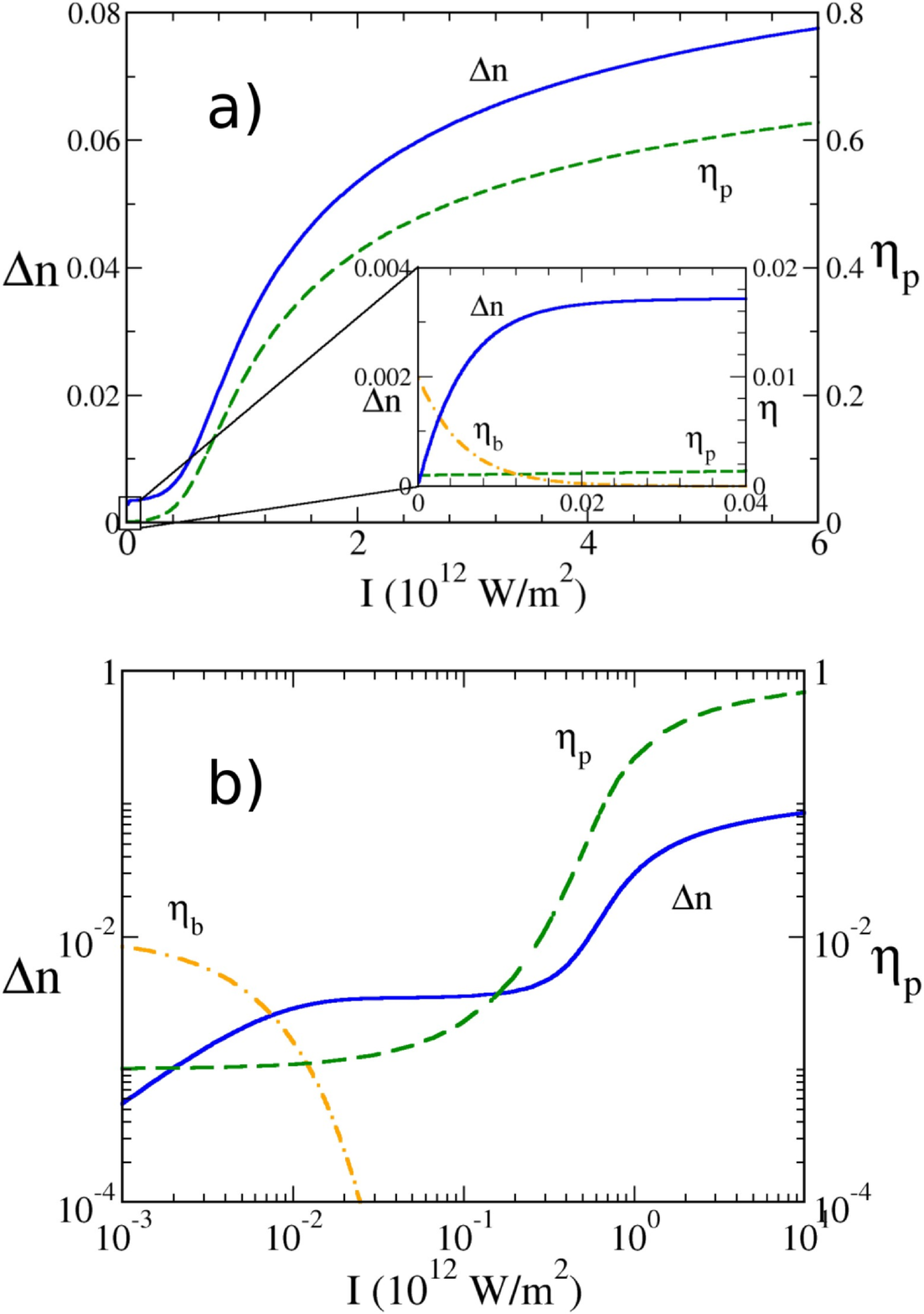}
\end{center}
\caption{(Color online) (a) Packing fractions of silica particles (dashed) and bubbles (dash-dotted), and the total induced refractive index change (solid)
vs.~the light intensity. The inset presents the dependence in the low intensities range. (b) The same presented in a bilogarithmic scale.
Thanks to the appropriate choice of particle sizes and concentrations (see text), 
the refractive index dependence has a form of two ``steps'', and supports two-dimensional soliton bistability. See text for values of parameters.
}\label{etau}
\end{figure}

The typical dependence of packing fractions of both components on the light intensity as well as the total induced refractive index change 
$\Delta n = n_{\rm eff} - n_0$
are presented in Fig.~\ref{etau}. The bubble packing fraction
is depleted almost completely already at relatively low light intensities due to the large bubble size and the high polarizability.
On the other hand, the concentration of silica particles remains low until the intensity reaches
a certain higher value. Further on, the concentration of the particles increases exponentially and finally saturates 
as the packing fraction becomes larger than $10\%$.
In result, thanks to the appropriate choice of particle sizes and concentrations, 
the refractive index dependence has a form of two ``steps''. 
The first step is a consequence of saturation of bubble induced nonlinear index change following bubble depletion~\cite{Segev},
while the second step is an effect of saturation of particle induced nonlinearity at high packing fractions due to limited available volume~\cite{colloidsOE}.
In the next section we show that this artificially prepared nonlinearity allows for the bistability of two-dimensional soliton beams.

\section{Two-dimensional solitons and soliton bistability}
\label{Sol}

\begin{figure}
\begin{center}
  \includegraphics[width=8.5cm]{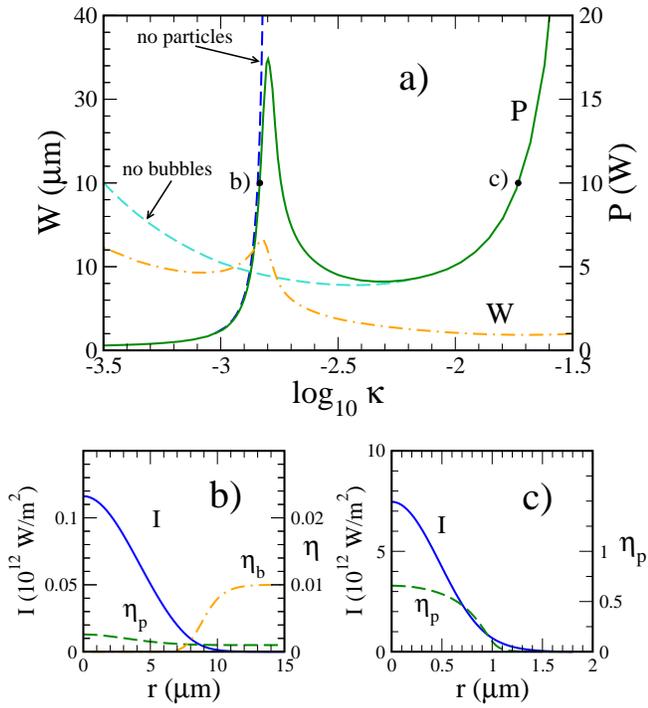}
\end{center}
\caption{(Color online)  (a) Soliton power (solid) and width (dash-dotted) vs.~the propagation constant $\kappa$. 
Two stable branches with ${\rm d} P / {\rm d} \kappa >0$ exist.
The dashed lines depict power dependence in the cases when one of the colloidal components (particles or bubbles)
is removed from the system.
Bottom panels show the soliton intensity profiles (solid) and colloidal particle (dashed) and bubble (dash-dotted) packing fractions
for two bistable solitons carrying power $P=10\,$W from (b) the lower stable branch and (c) the upper stable branch. In (c), the
bubble packing fraction is negligible. Notice the difference
in the width scale.} \label{Pkappa}
\end{figure}

We now consider two-dimensional spatial soliton solutions of
the attenuation-free version of Eq.~(\ref{PE}).
We look for localized solutions in the form of circularly symmetric beams $u({\bf r}) = A(r) \exp(i \kappa z)$ under the condition $\Gamma_\nu=0$. 
The propagation equation (\ref{PE}) reduces to
\begin{equation} \label{PEred}
-\kappa A + \frac{1}{2} \left(\frac{\partial^2 }{\partial r^2} + \frac{1}{r}\frac{\partial }{\partial r} \right) A + \sum_\nu \delta_\nu (\eta_\nu - \eta_{0\nu}) A = 0\,.
\end{equation}
The soliton profiles can be obtained using numerical relaxation methods, see eg.~\cite{Relax}.
The dependence of  the soliton power $P = \int |u|^2 \de \er$ and width $W = 3 \int r |u|^2 \de \er / P$ vs.~the propagation constant $\kappa$
is displayed in Fig.~\ref{Pkappa}(a). The branches of stable solutions correspond to positive slope ${\rm d} P / {\rm d} \kappa >0$ 
while unstable solutions are characterized by a negative slope~\cite{Kaplan}. The picture shows two stable branches, 
and within the power range $P\approx 4-17$~W stable solutions corresponding to both of the branches exist.
These bistable solitons fulfill all the three stability conditions required for their robustness during collisions~\cite{BistableCol}.

This picture should be compared with the results obtained in the case of a single colloidal component. 
The effect of soliton bistability was predicted in the one-dimensional case \cite{colloidsOE}, corresponding to a planar waveguide setup
or surface waves, however the range of powers supporting the bistability was significantly smaller. Moreover,
it was shown that in two-dimensional case the lower soliton branch becomes unstable \cite{PLP} due to the 
intrinsic instability of multidimensional solitons in Kerr media. The introduction of the second colloidal species is therefore
necessary for the bistability of two-dimensional solitons. Indeed, if one of the components is removed from the system,
only one of the stable branches remain, see dashed lines in Fig.~\ref{Pkappa}(a). It is clear that the bubbles determine the properties
of the system for low values of $\kappa$ and low light intensities, while the particles are the main acting component in the regime
of high $\kappa$. The bistability is the result of combination of the effects that these two species have on the nonlinear response
of the system.

In Figs.~\ref{Pkappa}(b,c) we present examples of bistable soliton profiles corresponding to the soliton power $P=10\,$W. 
It is clear that the light intensity of the lower branch soliton (b) corresponds to the first ``step'' from Fig.~\ref{etau},
while the upper branch soliton (c) corresponds to the second ``step''.
Hence, one can call the lower branch solitons ``bubble solitons'', while the upper branch solitons are ``particle solitons'', if referring to 
the main stabilizing component.
Since the width of the soliton from the lower branch is approximately $7\times$ larger than the width of the soliton from the upper branch carrying
the same beam power, these solitons can be easily distinguished in experiment. We note that
in the case of the lower branch solitons packing fractions
of both components are below 1\%, while the packing fraction of particles
is as high as 60\% in the center of the soliton beam from the upper branch. This value indicates that the simple
Carnahan-Starling model that we have used for description of particle interactions breaks down, and appearance
of ordered dense phase can be anticipated \cite{SimpleLiquids}. We note however that the breakdown of our model should not
lead to qualitative changes in the beam propagation and soliton properties, since the discrepancy in the particle concentration, 
which determines the nonlinear response, is 
relatively low. On the other hand, the packing fraction of bubbles is always below 1\% (the background packing fraction),
which confirms that our assumption of low bubble concentration is well justified.

\section{Soliton interactions}

\begin{figure}
\begin{center}
  \includegraphics[width=8.5cm]{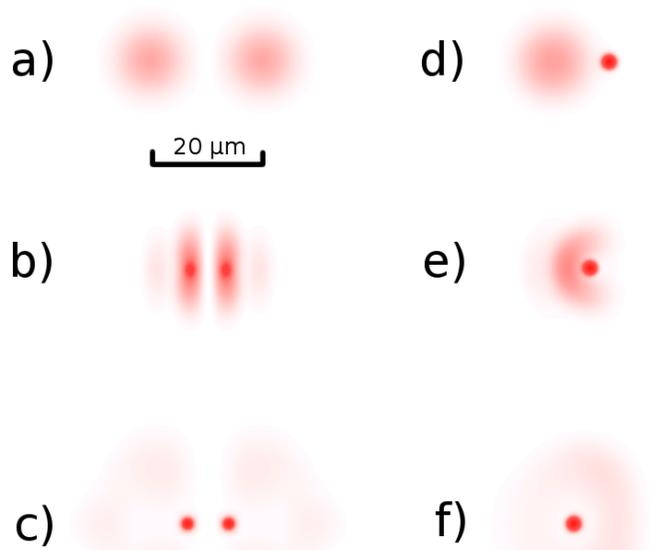}
\end{center}
\caption{(Color online) Collision of two identical out of phase solitons ($\Delta \varphi=\pi$) 
from the lower branch (left column) and of two solitons from different branches (right column). The power in each of the beams is
equal to $P=15\,$W (left) and $P=12\,$W (right), and the angle between the beams is $5^\circ$ and $2.5^\circ$, respectively.
The consecutive frames correspond to propagation distances $z=0$ (top row), $z=165\,\mu$m (middle row), and $z=425\,\mu$m (bottom row),
In the first case, 
repulsive interaction leads to switching of the solitons from the lower (a) to the upper branch (c).
In the second case, the interaction is also repulsive, independently of the relative phase between the solitons,
and leads to destruction of the wider but weaker soliton.  }\label{col}
\end{figure}

We proceed to the investigation of interactions of the solitons from the two bistable branches.
We consider a collision of two soliton beams angled towards each other which initially have the form of two stationary
soliton solutions, $u_1(x, y)$ and $u_2(x, y)$, separated by a distance $2x_0$ large in comparison to their widths.
The solitons have imprinted opposite linear phases $k_0$, which resemble the initial beam tilt, and a constant phase
difference $\Delta \varphi$
\begin{equation}
u(x,y,z=0) = u_1(x + x_0,y) \ee^{i k_0 x} + u_2(x - x_0,y) \ee^{-i k_0 x + i\Delta \varphi}\,.
\end{equation}
The initial profiles are taken as solutions to the undamped equation (\ref{PEred}), and the evolution of the beams is
modelled using the full equation (\ref{PE}) with the scattering losses included.

In Fig.~\ref{col} we present results for interaction of two identical out of phase solitons ($\Delta \varphi=\pi$) 
from the lower branch (a,b,c) and a collision of solitons from different branches (d,e,f). We find that,
similarly as in the one-dimensional case \cite{colloidsPRA}, interaction of two solitons from lower branch 
can lead to switching to the upper branch,  Fig.~\ref{col}(c). However, we were not able to observe a similar phenomenon
in the interaction of two solitons from different branches. Instead, the lower branch soliton is destructed
in most cases, see Fig.~\ref{col}(f). On the other hand, we find that the upper branch soliton is much more robust
and appears in the same form after the collision.
Nevertheless, we found that the interaction between solitons from different branches
is phase independent and always repulsive, in analogy with the one-dimensional case \cite{colloidsPRA}. 
This kind of interaction has been explained within the model of effectively incoherent beams due to the large difference
in propagation constant $\kappa$. Despite that each of the solitons appears as an attractive potential well for the
other soliton, repulsive interaction occurs if the collision angle is small enough \cite{colloidsPRA,defects}. To our knowledge, we
present the first example of soliton repulsion from attractive potential in two dimensions.

\section{Conclusions}

We have derived the model equations for description of a system of two colloidal components dispersed in a liquid medium in the presence of coherent laser light.
We have shown that with an appropriate choice of refractive indices and particle dimensions, it is possible to observe the phenomenon of optical soliton bistability 
in two spatial dimensions in a broad beam power range. 
Analogously as in the one-dimensional case, the interaction of soliton beams can lead to switching to the upper soliton branch,
and the interaction of solitons from different branches is phase-independent and always repulsive. 
The presented results can have implications for the experiments on optical solitons in 
soft matter media~\cite{Ashkin1,Gordon,Segev,Segev_OL,colloidsOE,Dholakia_OE,Asanto,Pispas,Likos}.

\section*{Acknowledgements}

This research was supported by the Australian Research Council and the Research School of Physics and Engineering of the Australian National University. 
The author would like to thank Wies\l{}aw Kr\'olikowski for many valuable discussions.

\clearpage

\end{document}